\documentclass[prl,twocolumn,superscriptaddress,showpacs,preprintnumbers]{revtex4-1}
\usepackage{amssymb}
\usepackage{graphicx}
\usepackage{dcolumn}
\usepackage{bm}
\usepackage{color}

\begin{document}

\title{Quantification of  correlations in quantum many-particle systems}
\author{Krzysztof Byczuk}
\affiliation{Physics Faculty, Institute of Theoretical Physics, University of  Warsaw,  Ho\.za 69, 00-681 Warszawa, Poland}
\author{Jan Kune\v{s}}
\affiliation{Institute of Physics, Academy of Sciences of the Czech Republic, Cukrovarnick\'a 10, 162 53 Praha 6, Czech Republic}
\author{Walter Hofstetter}
\affiliation{Institut f\"ur Theoretische Physik, Goethe-Universit\"at, 60438 Frankfurt/Main, Germany}
\author{Dieter Vollhardt}
\affiliation{Theoretical Physics III, Center for Electronic Correlations and Magnetism, Institute of Physics, University of Augsburg, 86135 Augsburg, Germany}
\date{\today }

\begin{abstract}
We introduce a well-defined and unbiased measure of the strength of correlations in quantum many-particle systems which is based on the relative von Neumann entropy computed from the density operator of correlated and uncorrelated states.  The usefulness of this general concept is demonstrated by quantifying correlations of interacting electrons in the Hubbard model and in a series of transition-metal oxides using dynamical mean-field theory.
\end{abstract}

\maketitle

Correlations in solids are the origin of many surprising phenomena such as Mott-insulating behavior and high-temperature superconductivity \cite{Mott90,Imada98,Tokura03}.
During the last few years the investigation of correlations between ultracold atoms in optical lattices has opened another fascinating field of research \cite{Bloch08}.

In many-body physics correlations are conventionally defined as the effects which go beyond factorization approximations such as Hartree-Fock theory \cite{Mahan}.
The actual strength of correlations in a given system is usually quantified by an interaction strength $U$ relative to an energy unit such as the bandwidth $W$, or by comparing expectation values of particular operators, e.g., the interaction or the total energy \cite{Grobe,Oles,Ziesche,Gottlieb,Harriman}.
However, there are many other quantities which can in principle be used to measure the correlation strength.
Indeed ``correlations'' are by definition a relative concept since they always require the comparison with some reference system.
 Any approach which employs the expectation value of a particular set of operators for comparison will be biased.
This raises a  fundamental question: Is there an  objective way to quantify the correlations of a system, which even allows one to compare  the correlation strength  of different systems?

Maximal information about a general quantum state is provided  by the corresponding density operator (``statistical operator'')
$\hat{\rho}=\sum_{i}p_i|\psi_i\rangle \langle \psi_i|$,
where $p_i$ is the probability for the quantum state $|\psi_i\rangle$ to be present in the mixture \cite{Peres}.
If there exists a basis in which $\hat{\rho}$  can be  completely factorized ($\hat{\rho}\rightarrow \hat{\rho}^{\rm PS}$, where PS refers to ``product state''), the corresponding state is by definition uncorrelated \cite{Horodecki09}.
To quantify the correlation strength of a quantum state one has to compare $\hat{\rho}$ with $\hat{\rho}^{\rm PS}$ in a suitable way.

In this Letter we propose to quantify correlations within a statistical approach which is based on  the concept of the von Neumann entropy \cite{Peres,Wehrl78}.
We will show that the relative entropy  \cite{Vedral02} of a quantum state with respect to  an uncorrelated product state  provides a well-defined, unbiased and useful measure of the correlation strength in that system.
This  concept not only allows one  to quantify correlations, but  even to compare the correlation strength of different quantum states.
In quantum  theory the von Neumann entropy is defined as $S (\hat{\rho}) = - \langle \ln \hat{\rho} \rangle_{\hat{\rho}}= -{\rm Tr} (\hat{\rho} \ln \hat{\rho})$.
It  quantifies the degree to which a quantum system is in a mixed state. In this way one can also define a relative entropy \cite{Vedral02}
\begin{eqnarray}
\Delta S(\hat{\rho}||\hat{\sigma})\equiv
  \langle \ln \hat{\rho} \rangle_{\hat{\rho}} - \langle \ln \hat{\sigma} \rangle_{\hat{\rho}}  ,
\label{relative-entropy}
\end{eqnarray}
provided the support
\cite{support} of
$\hat{\rho} $ is contained in the support of $ \hat{\sigma}$; otherwise
$\Delta S(\hat{\rho}||\hat{\sigma})$ is infinite.
By definition the relative entropy defined in this way is positive, monotonic, additive, and convex in its arguments \cite{appendix}.
The relative entropy measures ``how different'' two quantum states are \cite{Vedral02,Vedral97,Bjelakovic05,Schumacher,appendix}.
Therefore we propose to quantify correlations by the relative entropy $\Delta S(\hat{\rho}||\hat{\rho}^{\rm PS})$ between the statistical operator $\hat{\rho}$ of a system and an uncorrelated product state with $\hat{\rho}^{\rm PS}$ for  the same system \cite{appendix,com}.
Since  this approach uses a statistical operator it accounts for correlations represented by all possible correlation functions generated by $\hat{\rho}$. In the following we will make use of a reduced statistical operator which restricts the set of assessable correlation functions.

 The general approach works as follows \cite{Noack05,Rycerz06,Franca06,Larsson06,Amico}: Let $|\Psi\rangle$ be any state in the Hilbert space of the system.
We decompose the system into two subsystems yielding two Hilbert subspaces $A=\{|a\rangle\}$ and $B=\{|b\rangle\}$, where $|a\rangle$ and $|b\rangle$ are basis vectors  of these subspaces.
Then  one may write $|\Psi\rangle = \sum_{ab} \Psi_{a,b}|a\rangle |b\rangle$ with amplitudes $\Psi_{a,b}$.
We trace out the degrees of freedom of the B  subsystem.
The reduced statistical operator is given by $\hat{\rho}_A^{\Psi}=Tr_B |\Psi \rangle \langle \Psi | = \sum_{a_1, a_2}|a_1\rangle \rho_{a_1,a_2}^{\Psi}\langle a_2|$, with  the matrix elements $\rho_{a_1,a_2}^{\Psi}=\sum_{b} \Psi_{a_1,b} \Psi_{a_2,b}^{* }$.
By defining diagonal operators, i.e., projectors, $\hat{P}_{a}=|a\rangle \langle a|$, and  off-diagonal operators $\hat{T}_{a_1a_2}=|a_1\rangle \langle a_2|$, respectively, one can determine $\rho_{a_1,a_2}^{\Psi} =\langle\Psi| \hat{T}_{a_1, a_2} | \Psi \rangle^{\dagger}$ explicitly.

In the following  we will demonstrate the usefulness of this general concept by applying it to interacting lattice electrons described by the Hubbard model, i.e., a generic many-electron model with purely local interaction. In its simplest version it has the form
 \begin{eqnarray}
\hat{H}=\sum_{ij}t_{ij} \hat{c}^{\dagger}_{i\sigma}\hat{c} _{j\sigma} + U \sum_i \hat{n}_{i\uparrow} \hat{n}_{i \downarrow},
\label{hubbard}
\end{eqnarray}
where $\hat{c}^{\dagger}_{i\sigma}$ ($\hat{c} _{i\sigma}$) are creation (annihilation) operators for fermions of spin $\sigma=\uparrow(\downarrow)$ at the lattice site $i$,  $t_{ij}$ are hopping amplitudes between different sites, and $U$ is the local interaction energy.
The Hubbard model cannot be solved exactly.
As an approximation we therefore employ dynamical-mean field theory (DMFT), which becomes exact in the limit of infinite spatial dimensions ($d\rightarrow\infty$) \cite{Metzner89} and is known to provide a reliable local, but fully dynamical, description of correlated electron systems in $d=3$ \cite{Georges96,PT}.
By using  DMFT to compute the relative von Neumann entropy only the most important, namely local, correlations of the exact solution are taken into account.
In  DMFT the subspace $A$ is chosen to correspond to a site $i$ with the four state vectors $|a\rangle = \{|0\rangle, |\uparrow\rangle, |\downarrow\rangle,|2\rangle\}$ which  characterize the local occupation of the site.
By computing the statistical operator within  DMFT we determine the eigenvalues $p_{a}=\langle \Psi | \hat{P}_{a} |\Psi \rangle$ of the reduced statistical operator $\hat{\rho}_{\rm A}$.
We then employ an explicit product state $| {\rm PS} \rangle$ to calculate the uncorrelated reduced statistical operator $\hat{\rho}_{\rm A}^{\rm  PS}$ and its eigenvalues $p_a^{\rm PS}$. In this way the relative entropy (\ref{relative-entropy}) of the two states can be determined.
In the following, the subspace index $A \equiv i$ will be omitted.
In the absence of off-diagonal long-range order ($\langle\Psi | \hat{c}_{\sigma}|\Psi\rangle = 0$, $\langle\Psi |\hat{c}_{\sigma}\hat{c}_{-\sigma}|\Psi\rangle=0$) the reduced, i.e., local, statistical operator is diagonal \cite{Noack05,Rycerz06,Franca06,Larsson06,Amico,appendix}:
$ \hat{\rho} = p_0 |0\rangle \langle 0| + \sum_{\sigma} p_{\sigma}
|\sigma \rangle \langle \sigma| + p_2 |2\rangle \langle 2 |$.
The diagonal matrix elements $p_{a}$ are expressed by the particle number density $n$, magnetization density $m$, and double occupation per site $d$, i.e., $p_0=1-n+d$, $p_{\sigma}=(n+\sigma m)/2-d$,  and $p_2=d$.
Because of its diagonal form the local von Neumann entropy is given by $ S (\hat{\rho}) =  -\sum_{a} p_{a} \log p_{a}$ with $a=0,\; \uparrow, \downarrow,\; 2$. Then the local relative entropy becomes
\begin{eqnarray}
 \Delta S(\hat{\rho}||\hat{\rho}^{\rm PS})
 =  \sum_{a} p_{a} (\ln p_{a} - \ln p_{a}^{\rm PS}),
\label{reduced_relative_entropy}
\end{eqnarray}
where $ p_{a}^{\rm PS}$ is the corresponding expectation value  within the factorization approximation corresponding to $|\rm PS \rangle$.
 We note that the reversed relative entropy $ \Delta S(\hat{\rho}^{\rm PS}||\hat{\rho})$ is also defined.
The statistical operator is no longer diagonal if, for example, the local interaction includes nondensity type terms, or if superconducting phases are considered. In this case the statistical operator needs to be diagonalized to determine its logarithm.

To compute the parameters $n$, $m$, and $d$ the DMFT self-consistency equations are solved numerically by using the numerical renormalization group method at zero temperature \cite{NRG}.
We  assume  hopping $t_{ij}\equiv t$ between nearest neighbor sites $i,j$ on a  cubic lattice and use the corresponding density of states for the DMFT calculations.
The bandwidth $W=6t=1$ sets the energy unit. We fix the particle number density at $n=1$.
In the following both paramagnetic and antiferromagnetic ground states are considered.

\begin{figure}[tbp]
\includegraphics [clip,width=7.9cm,angle=-00]{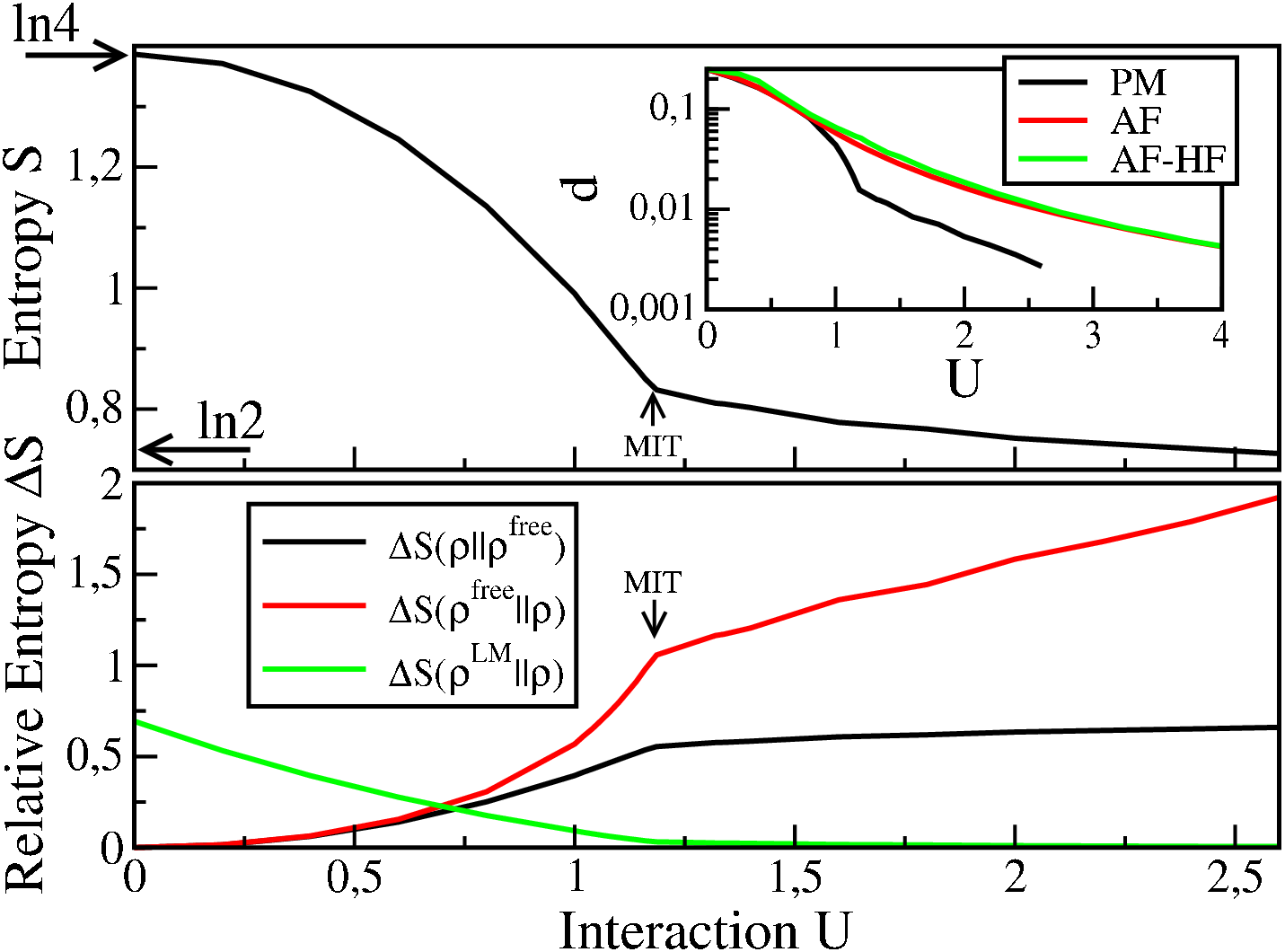}
\caption{
Upper panel: local von Neumann entropy of the correlated PM ground state as a function of the interaction $U$.
Inset: double occupation in the PM (black curve) and the AFM  (red curve) ground state
compared to the Hartree-Fock result (green curve) as a function of $U$.
Lower panel: local  relative entropies (PM phase) between the correlated ground state and the noninteracting ($U=0$) and local moment ($U=\infty$) ground states, respectively,  as functions of  $U$.
}
\label{fig1}
\end{figure}

(i) {\em Paramagnetic ground state}: We analyze the correlated paramagnetic (PM) ground state $|\rm   PM \rangle $ with respect to the following two reference states:  (i) the product  state in $k$ space $|{\rm PS }\rangle \equiv |{\rm free}\rangle$ corresponding to noninteracting  ($U=0$) electrons, where $d=(n^2-m^2)/4$,  and (ii) the product state in position space, i.e.,  the local moment state $|{\rm PS}\rangle\equiv |{\rm LM}\rangle$ where $d=0$.  In the PM phase $m=0$ implying that
$S(\hat{\rho}) =  -2\left[ d\ln d + (\frac{1}{2}-d)\ln (\frac{1}{2}-d)\right]$
is determined solely by the dependence of the double occupation $d$ on $U$.
In the upper panel of Fig.~\ref{fig1} the local entropy is shown as a function of $U$, and $d$ is shown in the  inset.
Both are monotonically decreasing functions of $U$: the local entropy decreases from $\ln 4$ at $U=0$ to $\ln 2$  at $U=\infty$ and the double occupation from $1/4$ to $0$. At half-filling a Mott-Hubbard metal insulator transition (MIT) takes place at a critical interaction $U_c\approx 1.225$, at which a correlation gap opens in the one-particle spectral function.
At the MIT the double occupation, and therefore the local entropy,  displays a kink.
The  free state $|{\rm  free}\rangle$ and the local moment state $|{\rm LM}\rangle $ have constant local entropies $S_{\rm free}= \ln 4$ and $S_{\rm LM}=\ln 2$, respectively, which reflect the degeneracy of those uncorrelated product states.

The correlated PM state $|\rm PM \rangle$ can be distinguished from the product states $|{\rm  free}\rangle$ and $|{\rm LM}\rangle $, respectively, by means of four different relative entropies: $\Delta S (\hat{\rho}||\hat{\rho}^{ \rm free})$, $ \Delta S (\hat{\rho}||\hat{\rho}^{ \rm LM})$, $\Delta S (\hat{\rho}^{ \rm free}||\hat{\rho})  $, and $\Delta S (\hat{\rho}^{ \rm LM}||\hat{\rho}) $, which are obtained from the three statistical operators $\hat{\rho}$, $\hat{\rho}^{\rm free}$, and $\hat{\rho}^{\rm LM}$ using Eq.~(\ref{reduced_relative_entropy}).
The relative entropies  are functions of $U$ as is shown in the lower panel of Fig.~\ref{fig1}.
The larger their value, the more distinguishable the two corresponding states  are.
In the Mott phase the relative entropy $\Delta S (\hat{\rho}||\hat{\rho}^{ \rm free})  $ increases much more slowly with the interaction strength $U$ than in the metallic phase.
This means that in a Mott insulator the correlation strength increases more slowly with $U$ than in a correlated  metal.
We note that  $\Delta S (\hat{\rho}||\hat{\rho}^{ \rm LM}) =\infty$ for $U<\infty$  because in the local moment state double occupation is completely excluded ($p_2^{\rm LM}=0$), i.e.,  $|{\rm PM}\rangle$ and $|{\rm LM}\rangle$ can be perfectly distinguished by measuring the respective double occupation of the two states. 
By contrast, the reverse relative entropy $\Delta S (\hat{\rho}^{ \rm LM}||\hat{\rho})$ remains finite although the two states are perfectly distinguishable. This shows that $\Delta S (\hat{\rho}||\hat{\rho}^{ \rm LM})$ rather than the reverse expression is the more appropriate measure of the correlations.

\begin{figure}[tbp]
\includegraphics [clip,width=7.9cm,angle=-00]{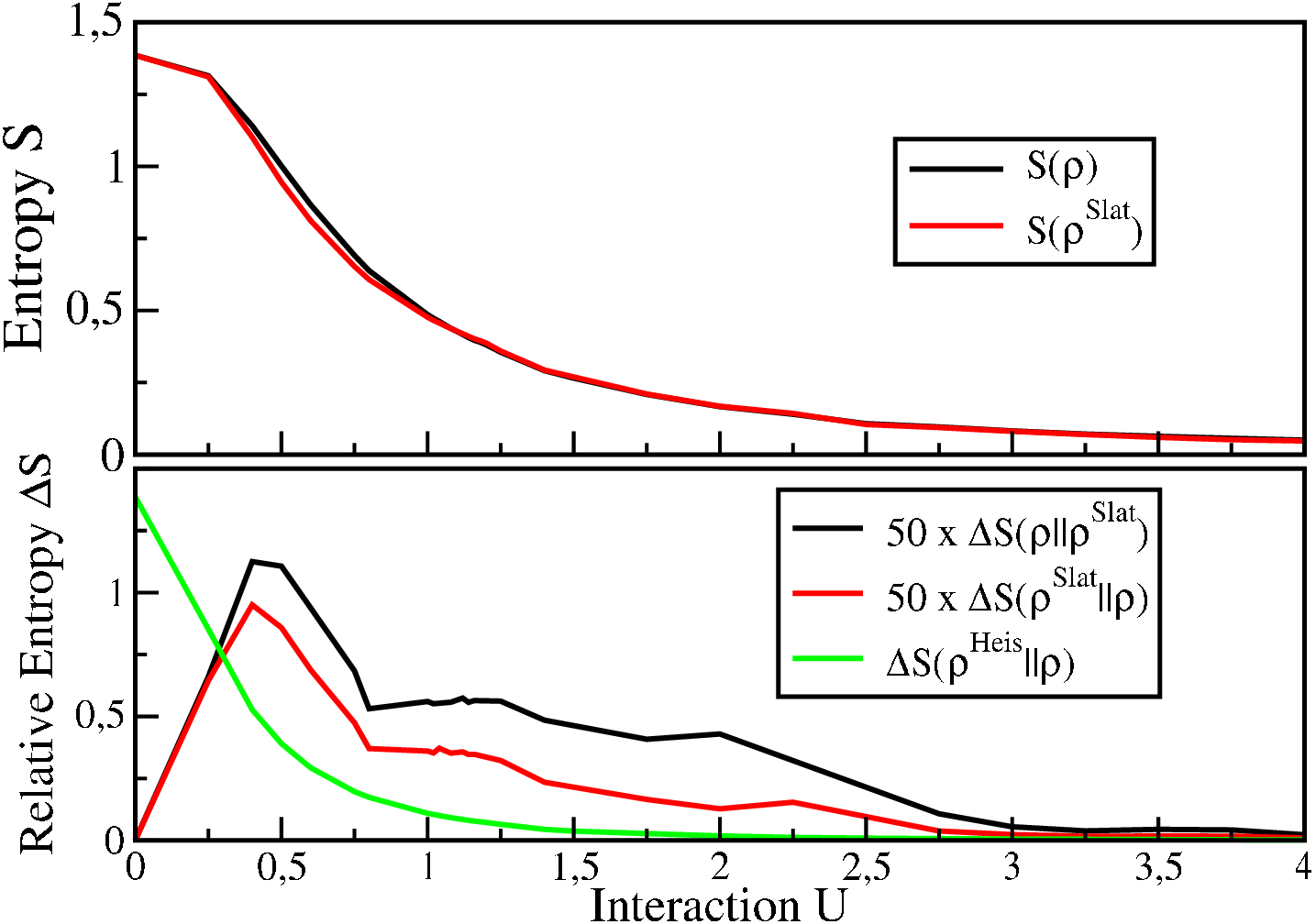}
\caption{Upper panel:  local von Neumann entropy $S$ of the correlated AFM and the Slater ground states as a function of $U$.
Lower panel: local relative entropies $\Delta S$ between correlated AFM, Slater, and Heisenberg  ground states.  $\Delta S(\hat{\rho}||\hat{\rho}^{ \rm Slat}) $ and $ \Delta S(\hat{\rho}^{ \rm Slat} ||\hat{\rho} ) $ are magnified by a factor of $50$.
}
\label{fig2}
\end{figure}

(ii) {\em Antiferromagnetic case}: Next we compare the correlated antiferromagnetic (AFM) ground state $|{\rm AF}\rangle$ with the product states represented by the Slater state $|{\rm   Slat}\rangle$ (a spin-density product state in ${k}$ space) and the Heisenberg state $|{\rm   Heis}\rangle$ (a perfect Ne\'el-type  product state in position space), respectively.
 The staggered magnetization $m_{\rm st}$ of the Slater state is determined within  Hartree-Fock, whence $d_{\rm Slat}=((n_{\rm Slat})^2-(m_{\rm   st,Slat})^2)/4$. In the Heisenberg state $m_{\rm st,Heis}=1$ and $d_{\rm   Heis}=0$.
As  shown in the upper panel in Fig.~\ref{fig2} the local entropies $S(\hat{\rho})$ and $S(\hat{\rho}^{\rm Slat})$ are almost identical.
A similar agreement is found when the local entropies are plotted as functions of the order parameter $m_{\rm st}$ instead of the interaction $U$.
Both local entropies decrease from $\ln 4$ at $U=0$ to zero at $U=\infty$. This is due to the presence of  long-range order with a magnetization that fully saturates.
 It means that  at $U=\infty$ the spin degeneracy is completely lifted. We observe that $ S(\hat{\rho}) > S(\hat{\rho}^{\rm Slat}) $ because $m<m_{\rm Slat}$ and $d<d_{\rm Slat}$ due to quantum fluctuations, cf.  inset of Fig.~\ref{fig1}.
The Slater (weak coupling) and the Heisenberg (strong coupling) limits are smoothly connected \cite{byczuk09}.

The correlated AFM state $|{\rm AF}\rangle$ can be distinguished from the product states $|{\rm  Slat}\rangle$ and $|{\rm Heis}\rangle $, respectively, by the four relative entropies $\Delta S(\hat{\rho}||\hat{\rho}^{ \rm Slat})$, $ \Delta S(\hat{\rho}||\hat{\rho}^{ \rm Heis} )$, $ \Delta S(\hat{\rho}^{ \rm Slat} ||\hat{\rho} ) $, and $ \Delta S(\hat{\rho}^{ \rm Heis}||\hat{\rho} ) $ which are obtained from the three statistical operators $\hat{\rho}$, $\hat{\rho}^{\rm Slat}$, and $\hat{\rho}^{\rm Heis}$.
The relative entropies $ \Delta S(\hat{\rho}||\hat{\rho}^{ \rm Slat})$ and $ \Delta S(\hat{\rho}^{ \rm Slat} ||\hat{\rho} ) $ are more than one order of magnitude smaller than in the PM case, as is seen in the inset to the upper panel of Fig.~\ref{fig2}.
As before the relative entropy $  \Delta S(\hat{\rho}||\hat{\rho}^{ \rm Heis} ) =\infty$  because $d_{\rm Heis}=0$, implying $p_2^{\rm Heis}=0$ for any $U$.
By contrast, $  \Delta S(\hat{\rho}^{ \rm Heis}||\hat{\rho} )  $ is finite and approaches zero at large $U$.
Correlationwise the Heisenberg product state and the correlated AFM state are therefore very similar at large $U$, and cannot be easily  distinguished.
This means that,  in contrast to the PM state, the AFM correlated state is locally only weakly correlated.

\begin{figure}[tbp]
\includegraphics [clip,width=7.9cm,angle=-00]{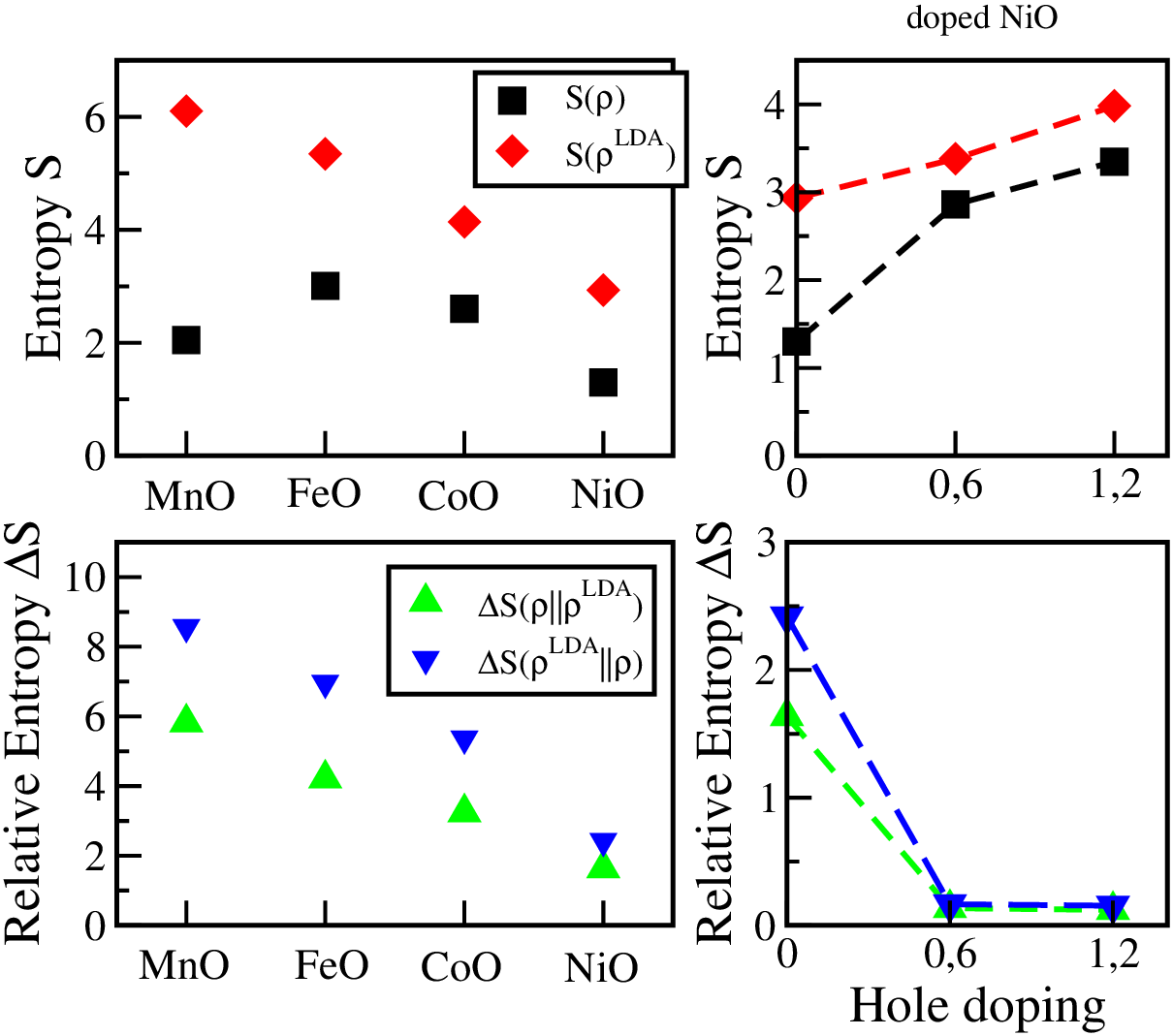}
\caption{Left panels: Local von Neumann (top) and relative  (bottom) entropies for MnO, FeO, CoO, and NiO series.
Right panels: Local von Neumann (top) and relative  (bottom) entropies for hole doped NiO; see text.
} \label{fig3}
\end{figure}

In a further application, we employ the relative entropy to quantify correlations in real materials.
As an example we select a series of transition-metal (TM)  monoxides MnO, FeO, CoO, and NiO.
The results are obtained within the LDA+DMFT approach \cite{kotliar06,mno-kunes,nio-kunes} which combines the local density approximation (LDA) to density functional theory with the DMFT \cite{appendix}.
We take the states obtained from  the LDA  $|{\rm LDA} \rangle $ as the uncorrelated states.
In Fig. \ref{fig3} we show the local  entropies of the series of TM monoxides.
The local entropies computed by LDA+DMFT describe the corresponding ionic ground states, which correspond in all four cases to high-spin configurations.
The orbital degrees of freedom imply that NiO and MnO have a twofold degeneracy, while FeO and CoO have a sixfold degeneracy \cite{note}.
The trend in the LDA  entropy reflects the decreasing number of possibilities to distribute 5 (MnO), 6 (FeO), 7 (CoO), and 8 (NiO) electrons among 10 orbitals, which are  modified by the crystal-field splitting and deviate from integer occupation.
The dominant effect of the  correlations is to reduce the large number of local many-body ground states available in the $|\rm LDA\rangle $ solution to only a few in the $|\rm DMFT \rangle$ solution.
We conclude that MnO, with a  $d^5$ configuration, is more correlated than NiO which has a $d^8$ configuration, because in a $d^5$ configuration the interaction modifies the local occupation matrix more significantly.

Finally, we investigate the effect of hole doping on NiO  \cite{nio-kunes}.
Reducing the $3d$ occupation towards half-filling increases the number of local states with non-negligible contribution to the ground state and leads to  an increase of the local entropy with hole doping  in the uncorrelated system.
In the correlated system the local density matrix changes substantially.
While in  stoichiometric NiO the $d^8$ ($t_{2g}^6e_g^2$) states dominate the local density matrix,  hole doping leads to a progressive population of $d^7$ and $d^9$ ($t_{2g}^6e_g^{1,3}$), respectively.
The increase of the number of states which effectively contribute to the ground state is reflected in the increase of the local entropy.
This leads to a lowering of the relative entropy with the number of holes and to a substantial decrease of correlations.

In conclusion, we employed the relative von Neumann entropy between correlated and factorized (uncorrelated) statistical operators to quantify  the correlation strength in quantum many-body systems.
We demonstrated the usefulness of this approach by  computing the degree of correlation in the Hubbard model and in a series of transition monoxides using DMFT and LDA$+$DMFT, respectively.
It will be interesting to apply this general concept to investigate correlations beyond a single-site approximation, in which case not only dynamical local, but also spatial correlations are included.

After publication of  the paper we learned that the relative entropy was
already proposed by A. D. Gottlieb and N. J.Mauser \cite{Mauser07} as a means to
quantify   correlations of a many-fermion state by comparing its density operator
  with that of the corresponding free state.

We thank R. Demkowicz-Dobrza\'nski, K. Held, A. Rycerz, and G. Uhrig for useful discussions.  This work was supported in part by the Deutsche Forschungsgemeinschaft through Transregio TRR 80 (KB, DV), Transregio TRR 49 (WH), Forschergruppe FOR 1346 (JK), and Forschergruppe FOR 801 (WH). One of us (KB) also acknowledges support by the Foundation for Polish Science (FNP) through the TEAM/2010-6/2 project cofinanced by the EU European Regional Development Fund.

%%%%%%%%%%%%%%%%%%%%%%%%%%%%%%%%%%%%%%%%%%%%%%%

%\end{document}

\vspace{0.5cm}

%\newpage

{\bf \Large Appendix }\\

\vspace{0.25cm}

{\bf A. Probabilistic interpretation of the relative entropy}\\

\vspace{0.25cm}

In quantum  theory the von Neumann entropy is given by the expectation value of $\log \hat{\rho}$, i.e., $S (\hat{\rho}) = - \langle \log \hat{\rho} \rangle_{\hat{\rho}}= -{\rm Tr} \hat{\rho}
\log \hat{\rho}$  quantifies the degree to which a quantum system is in a
mixed state.
In this way one can define a \emph{relative entropy} \cite{Vedral02}
\begin{eqnarray}
\Delta S(\hat{\rho}||\hat{\sigma})\equiv 
\langle \log \hat{\rho} \rangle_{\hat{\rho}} - \langle \log \hat{\sigma} \rangle_{\hat{\rho}}  \nonumber \\
= {\rm Tr}( \hat{\rho} \log
\hat{\rho} - \hat{\rho} \log \hat{\sigma}),
\label{relative-entropy}
\end{eqnarray}
provided the support
\cite{support} of
$\hat{\rho} $ is contained in the support of $ \hat{\sigma}$ (otherwise
$\Delta(\hat{\rho}||\hat{\sigma})$ is infinite).
The relative entropy measures ``how different'' two quantum states are \cite{Vedral02,Vedral97,Bjelakovic05,Schumacher}.
Let us assume that we experimentally investigate a system in an (unknown) state
with a statistical operator $\hat{\sigma}$.
We wish to determine whether the system is in a particular \emph{known} state with a statistical operator
$\hat{\rho}$, or not. While for the measurement on a single system
the answer can only be yes or no, an experiment on $N \gg 1$ many copies of the system
% in the
%state with statistical operator $\hat{\sigma}$
allows one to make a probabilistic statement about whether
the conclusion drawn from this measurement is right or wrong.
%For an ensemble with $N \gg 1$ copies
The probability for being able to
distinguish a system in the unknown state with statistical operator $\hat{\sigma}$ from a system in a
known  state with statistical operator $\hat{\rho}$ is  given by $P(\hat{\rho}|\hat{\sigma}) = 1-\exp
[ -N  \Delta S(\hat{\rho}||\hat{\sigma})]$.
If the states are the same, in which case $\hat{\sigma}=\hat{\rho}$, then $\Delta S(\hat{\rho}||\hat{\rho})=0$,
i.e., $P(\hat{\rho}|\hat{\rho}) =0$, implying that the two systems cannot be
distinguished.
In the another extreme limit, when the support of $\hat{\rho} $ is not
included in  the support of $ \hat{\sigma}$, the probability of discriminating between
the state with $\hat{\sigma}$ and the state with $\hat{\rho}$ is unity, implying that
the two systems can be distinguished with certainty.
Altogether the relative entropy
describes how likely it is for two quantum states to be distinguishable.
%In this Letter we propose to employ
The  relative entropy $\Delta S(\hat{\rho}||\hat{\rho}^{\rm PS})$ between a correlated system described by its  statistical operator $\hat{\rho}$ and an uncorrelated product state with $\hat{\rho}^{\rm PS}$ is used here to quantify correlations in this system. Thereby it is not only possible to \emph{quantify correlations in a system}, but also to \emph{compare} the degree of correlation of different systems. In the paper  we provide several examples which demonstrate the usefulness of this new concept of quantifying correlation in particular systems.

\vspace{0.25cm}

{\bf B. Propertes of the relative entropy and correlation measure}\\

\vspace{0.25cm}

 From the properties of the relative entropy it follows that: i) for an uncorrelated state with density operator $\hat{\rho}^{\rm PS}$ the relative entropy (\ref{reduced_relative_entropy}) is zero, since  $ \Delta S(\hat{\sigma}||\hat{\sigma})=0$ for any density operator $\hat{\sigma}$, ii) mixing a correlated and a product state decreases the  correlation strength  $\Delta S(p \hat{\rho}+(1-p)\hat{\rho}^{\rm PS} ||\hat{\rho}^{\rm PS}) \leq p \Delta S(\hat{\rho}||\hat{\rho}^{\rm PS}) + (1-p) \Delta S(\hat{\rho}^{\rm PS}||\hat{\rho}^{\rm PS}) =  p \Delta S(\hat{\rho}||\hat{\rho}^{\rm PS}) $, with $0\leq p \leq 1$, and iii) increasing the number of states increases the correlation strength, i.e., $ \Delta S(\otimes_i \hat{\rho}_i||\otimes_i\hat{\rho}^{\rm PS})=\sum_i  \Delta S(\hat{\rho}_i||\hat{\rho}^{\rm PS}_i)$. \\

\vspace{0.25cm}

{\bf C. Computation of the diagonal elements of the statistical operator}\\

\vspace{0.25cm}

In the atomic representation the local diagonal and off-diagonal operators (``Hubbard operators'') required to compute the matrix elements $\rho_{a_1,a_2 }^{\Psi}$ take the form:
\begin{eqnarray}
\hat{P}_{a}=\left\{
\begin{array}{c}
(1-\hat{n}_{\uparrow})(1-\hat{n}_{\downarrow})\\
\hat{n}_{\uparrow}(1-\hat{n}_{\downarrow})\\
(1-\hat{n}_{\uparrow})\hat{n}_{\downarrow}\\
\hat{n}_{\uparrow}\hat{n}_{\downarrow},
\end{array}
\right. \nonumber
\end{eqnarray}
and
\begin{eqnarray}
\hat{\tilde{T}}_{a_1 a_2}=\left(
\begin{array}{cccc}
1 & c_{\uparrow} & c_{\downarrow} & c_{\downarrow} c_{\uparrow}   \\
c_{\uparrow}^{\dagger} & 1 & c_{\uparrow}^{\dagger}  c_{\downarrow} & - c_{\downarrow}\\
c_{\downarrow}^{\dagger} & c_{\downarrow}^{\dagger}c_{\uparrow} & 1 & c_{\uparrow}\\
c_{\uparrow}^{\dagger}  c_{\downarrow}^{\dagger} & - c_{\downarrow}^{\dagger} & c_{\uparrow}^{\dagger}& 1
\end{array}
\right). \nonumber
\end{eqnarray}
Then $\rho_{a_1,a_2}^{\Psi} =\langle\Psi| \hat{T}_{a_1, a_2} | \Psi \rangle^{\dagger} = \langle\Psi| P_{a_1} \hat{\tilde{T}}_{a_1, a_2} P_{a_2} | \Psi \rangle^{\dagger}$.

\vspace{0.25cm}

{\bf D. The  LDA+DMFT computational scheme}\\

The Hamiltonians spanning the Hilbert space of TM-$3d$ and O-$2p$ bands in the Wannier representation were obtained using the wien2k and wannier90  codes [A1].
The on-site interaction, approximated by a  density-density form \cite{nio-kunes}, is diagonal in the occupation number basis, and was assumed to reside on the TM sites.
This, together with the fact that the non-interacting local propagator is diagonal in the basis of cubic harmonics ($e_g$--$t_{2g}$), leads to great simplification of the local occupation matrix.
Namely,  it is then diagonal in the occupation number basis.
Its diagonal elements are simply obtained by measuring separately the contributions of all $2^{10}$ states to the local trace with a continuous-time quantum Monte-Carlo simulation [A2].
The corresponding statistical operators are diagonal $\hat{\rho}=\sum_d p_d |p_d\rangle \langle p_d|$, where $\{|p_d \rangle \}$ is the $d$-atomic orbital base.
Therefore, all local entropies can be  easily determined. \\

\vspace{0.25cm}

[A1] Cf., A. A. Mostofi, J. R. Yates, Y.-S. Lee, I. Souza, D. Vanderbilt and N. Marzari
Comput. Phys. Commun. {\bf 178}, 685 (2008); J. Kunes, R. Arita, Ph. Wissgott, A. Toschi, H. Ikeda, K. Held, arXiv:1004.3934.\\

\vspace{0.25cm}

[A2] E. Gull, A.J. Millis, A.I. Lichtenstein, A.N. Rubtsov, M. Troyer, and Ph. Werner, Rev. Mod. Phys. {\bf 83}, 349 (2011).


\begin{references}
\bibitem{support} The support of a statistical operator $\hat{\rho}$ is defined as the subspace of a Hilbert space spanned by the eigenvectors corresponding to nonzero eigenvalues of $\hat{\rho}$.
\bibitem{com} In general, a state with the density operator $\hat{\rho}$
might contain both classical correlations and quantum correlations
(entanglement).  To extract unique information on the entanglement in
the states with $\hat{\rho}$ one should find that state $\hat{\rho}^{\rm
PS}$ for which the relative entropy is minimal \cite{Vedral02}. In our
paper this goal is not pursued since
the quantification of classical and quantum correlations in
many-particle systems, which are in a mixed state and in the
thermodynamic limit, is a hard computational problem \cite{Vedral02,Gharibian}. 
For the same reason we also do not search for product states by variationally optimizing the ground state energy. Instead we quantify correlations with respect to physically well-known uncorrelated states.
\bibitem{note} We note that the degeneracy two in the case of NiO and MnO is an 
artifact of our approximate interaction term. With the full Coulomb 
interaction the degeneracies would be three and six, respectively. 
Similarly the degeneracy of six in the case of FeO and CoO would change 
to 15 and 12, respectively, when the full Coulomb interaction is taken 
into account.
\end{references}
\end{document}